\documentclass[aps,preprint,showpacs]{revtex4}
\usepackage{graphicx}
\usepackage{bm}
\usepackage{dcolumn}
\usepackage{bm}

\begin{document}

\title{Extension of the Nambu--Jona-Lasinio model predictions at high densities and
temperatures by using an implicit regularization scheme}
\author{R.L.S. Farias$^1$, G. Dallabona$^2$, G. Krein$^3$, and O.A. Battistel$^4$}

\affiliation{$^1$Departamento de F\'\i sica Te\'orica, Universidade do Estado do Rio de
Janeiro, Rio de Janeiro, RJ 20559-900, Brazil}

\affiliation{$^2$Departamento de Ci\^{e}ncias Exatas, Universidade Federal de Lavras,\\
Cx. Postal 37, 37200-000, Lavras, MG, Brazil}

\affiliation{$^3$Instituto de F\'{\i}sica Te\'orica, Universidade Estadual 
Paulista, Rua Pamplona 145, S\~ao Paulo, SP 01405-900, Brazil}

\affiliation{$^4$Departamento de F\'{\i}sica, Universidade Federal de Santa Maria,
97119-900 Santa Maria, RS, Brazil}

\begin{abstract}
Traditional cutoff regularization schemes of the Nambu--Jona-Lasinio model
limit the applicability of the model to energy-momentum scales much below
the value of the regularizing cutoff. In particular, the model cannot be
used to study quark matter with Fermi momenta larger than the cutoff. In the
present work an extension of the model to high temperatures and densities
recently proposed by Casalbuoni, Gatto, Nardulli, and Ruggieri is used in
connection with an implicit regularization scheme. This is done by making
use of scaling relations of the divergent one-loop integrals that relate
these integrals at different energy-momentum scales. Fixing the pion decay
constant at the chiral symmetry breaking scale in the vacuum, the scaling
relations predict a running coupling constant that decreases as the
regularization scale increases, implementing in a schematic way the property
of asymptotic freedom of quantum chromodynamics. If the regularization scale
is allowed to increase with density and temperature, the coupling will
decrease with density and temperature, extending in this way the
applicability of the model to high densities and temperatures. These results
are obtained without specifying an explicit regularization. As an
illustration of the formalism, numerical results are obtained for the finite
density and finite temperature quark condensate, and to the problem of color
superconductivity at high quark densities and finite temperature.
\end{abstract}


\maketitle

\section{Introduction}

\label{sec:1}

The Nambu--Jona-Lasinio (NJL)~\cite{NJL} model has been the prototype model
for studying chiral symmetry restoration in hadronic matter at finite baryon
densities $\rho_B$ and finite temperatures $T$. Since the earlier
applications of the model at high $\rho_B$ and $T$~\cite{HK1,BMZ}, an
extensive and important body of work has been done in this direction -- for
reviews and a comprehensive list of references see Refs.~\cite%
{VW,Kl,HK-rev,Bc,Bj,Bpt,Rk}. Because the model is nonrenormalizable, the
high momentum part of the model has to be regularized in a phenomenological
way. The common practice has been to regularize the divergent loop
amplitudes with a three-dimensional momentum cutoff $\Lambda \sim 1$~GeV,
which also sets the energy-momentum scale for the validity of the model.
That is, the model cannot be used for studying phenomena involving momenta
running in loops larger than $\Lambda$. In particular, the model cannot be
used to study quark matter at high densities $\rho_B \sim k^3_F$ with $k_F >
\Lambda$, where $k_F$ is the quark Fermi momentum. One of the symptoms of
this problem is the prediction of vanishing superconducting gaps at high
baryon densities, a feature of the model that is solely caused by
the use of a regularizing momentum cutoff $\Lambda$ of the divergent
loop integrals~\cite{Casal,ja,skp}.

Recently, we~\cite{FDKB} have presented an alternative to the cutoff
regularization of the NJL model, in which the one-loop integrals are
reorganized through mathematical identities in a way finite integrals
become separated from density-independent divergent integrals. The
finite integrals are integrated without imposing any restriction on
the integration momenta and the divergent integrals are related to
physical quantities at the dynamical chiral symmetry breaking in
vacuum. As a result, instead of a vanishing gap like in the cutoff
regularization one obtains a finite gap that grows with density.
However, this does not mean that the model can be used at arbitrarily
high densities. The point is that any
implicit regularized integral contains an implicit scale, the
regularization scale. When such an integral is fitted to a
physical quantity in vacuum, we are implicitly also fitting this
scale. Now, in a loop integral at nonzero density and/or
temperature, the vacuum part is determined by the regularization
scale and the finite integrals are governed by the chemical potential 
and temperature. When these scales are very different,
one obtains nonsense results. Nevertheless, this mismatch of scales
is less problematic in the implicit regularization scheme than
in any other cutoff regularization scheme, in the sense that one
can go to much higher densities and temperatures before the problem
shows up.

The root of the problem is of course the nonrenormalizability of
the model. In a renormalizable theory all the dependence on the
cutoff can be removed in favor of running physical parameters,
like the coupling constants of QED and QCD. The running is given
by the renormalization group equations of the theory and is
controlled by an energy scale that is adjusted to the scale
of the experimental conditions under consideration. In a recent
publication, Casalbuoni et al.~\cite{Casal} have introduced the
concept of a running coupling constant for the NJL model to
extend the applicability of the model to high density. The
arguments are based on making the cutoff density dependent,
using an analogy with an ordinary solid. In an ordinary solid,
there is a natural maximum phonon frequency - the Debye
frequency - that increases as the lattice spacing gets smaller.
In quark matter, as the density of quarks
increases, the quarks get closer together and therefore the ultraviolet
cutoff $\Lambda $ of the NJL model should be allowed to increase
correspondingly. The cutoff can be changed consistently without spoiling the
predictions of the model for the chiral properties of the vacuum if the
four-fermion coupling constant $G$ of the model is allowed to change with
the cutoff. In Ref.~\cite{Casal}, the cutoff dependence of $G$ is obtained
from the joint consideration of the divergent expressions for the pion decay
constant $f_{\pi }$ and the gap equation for the constituent quark mass $M$:
fixing the value of $f_{\pi }$ at $93$~MeV, the divergent expression of $
f_{\pi }$ (in the chiral limit), regulated with a cutoff $\Lambda $, leads
to a constituent quark mass $M$ that is $\Lambda $ dependent. The use of this $
M=M(\Lambda )$ into the gap equation gives rise to a coupling $G$ that runs
with $\Lambda $. Moreover, the running of the coupling is such that $
G(\Lambda )$ decreases with increasing $\Lambda $. This is certainly
physically motivated and is also in accord with the interpretation~\cite
{HK1,BJM} that the cutoff in the NJL model simulates - albeit in a crude way
- the property of asymptotic freedom of QCD, in the sense that the coupling
between quarks decreases as higher momentum scales are probed. A similar
suggestion was made by Shakin and collaborators~\cite{Shk} in studies of
hadronic correlation functions. They found that results in qualitative
accord with lattice QCD calculations can be obtained once a significant
temperature dependence of the cutoff and of the coupling $G$ is allowed.

The aim of the present paper is to show that it is possible to extend the
applicability of the NJL model to high densities and temperatures without
the use of an explicit regularization of divergences. The basic motivation
for avoiding an explicit regularization such as those commonly used like the
three- or four-momentum cutoff, Pauli-Villars and proper-time
regularizations, is that these lead to global and gauge symmetry violations,
and to the breaking of causality and unitarity. Although in many situations
these problems do not have great influences on the final numerical results,
there are situations, however, where they do have drastic consequences, like
in studies of correlation functions. Arguments have been used and tricks
invented to circumvent such problems, but no satisfactory solution has been
found -- for a good discussion on these issues, see for example Refs.~\cite
{Bron,Doring,DavERA}.

The implicit regularization scheme has been originally proposed in
Ref.~\cite{thesis} and was used in different contexts~ \cite{impl-diff,Nemes},
including applications to the NJL model~\cite{BN,BK,FDKB,BDK}. At the
one-loop approximation, there appears only two divergent integrals, one
is quadratically divergent and the other is logarithmically divergent.
Once the divergent integrals are assumed to be
implicitly regulated, i.e. regulated without the specification an explicit
regulator, they will depend implicitly on a momentum scale that we denote by
$\Lambda $. These integrals satisfy well-defined scaling relations, in that
the integrals at some mass scale can be related to a combination of the same
integrals at some another, arbitrary scale. These scaling relations allow
to express the divergent parts of the amplitudes at finite temperature
and density in terms of their counterparts at zero temperature and density.
Moreover, fixing $f_{\pi }$ in the vacuum, one can derive a scaling relation
for the four-fermion coupling $G$ which gives a running $G(\Lambda )$
similar to that found in Ref.~\cite{Casal}. All this is achieved, we
reiterate, without specifying an explicit regularization scheme.

In the next section we review the main aspects of the implicit
regularization scheme~\cite{thesis,impl-diff,BN,BK,FDKB,BDK} that are most
relevant for the present paper. We start specifying the Lagrangian density
we are going to use and then explain the scaling relations and their use for
obtaining the solution of the gap equation within this scheme. In order to
make contact with results of the literature, we solve numerically this gap
equation at finite temperature and density and show the results for the
quark condensate as a function of temperature $T$ and quark chemical
potential $\mu $. In Section~\ref{sec:running}, we make use of the scaling
relations to obtain the running of the coupling $G$ of the model and compare
results with Ref.~\cite{Casal}. In Section~\ref{sec:CSC} we illustrate the
use of our results for obtaining the critical temperature as a function of
quark chemical potential for the spin-$0$ two-flavor color superconducting
(2SC) gap. Our conclusions and outlook are presented in Section~\ref
{sec:concl}.

\section{Calculational strategy to handle the divergences}

For the purposes of the present work, it is sufficient to consider the
simplest version of the model specified by the two-flavor $SU(2)$
Lagrangian density
\begin{equation}
{\mathcal{L}}_{NJL}=\bar{\psi}(i\!\not{\!}\partial -m_{0})\psi +G[(\bar{\psi}
\psi )^{2}-(\bar{\psi}\gamma ^{5}\vec{\tau}\psi )^{2}].
\end{equation}
Here $\psi $ is the quark field operator (with color and flavor indices
suppressed), $m_{0}$ is the current-quark mass matrix. At the one-loop
approximation, the gap equation at finite temperature $T$ and quark chemical
potential $\mu $, is given by (for simplicity we work in the chiral limit $
m_{0}=0$)
\begin{equation}
M=48\,G\,M\,T\sum_{n}\int \frac{d^{3}k}{(2\pi )^{3}}\;\frac{1}{(i\omega
_{n}+\mu )^{2}+E(k)^{2}},
\end{equation}
with $E(k)=\sqrt{k^{2}+{M}^{2}}$, and $\omega _{n}=(2n+1)\pi T$, $n=0,\pm
1,\pm 2,\cdots $ being the Matsubara frequencies. Performing the sum, one
obtains
\begin{equation}
M=48\,G\,M\left[ iI_{quad}(M)-I(T,\mu )\right] ,
\label{Gap_M}
\end{equation}
where
\begin{equation}
I_{quad}({M}^{2})=\int \frac{d^{4}k}{(2\pi )^{4}}\frac{1}{k^{2}-{M}^{2}} ,
\label{Iq-div}
\end{equation}
and
\begin{equation}
I(T,\mu )=\int \frac{d^{3}k}{(2\pi )^{3}}\;\frac{\left[ n_{-}(k)+n_{+}(k)
\right] }{2\,E(k)} ,
\end{equation}
with $n_{\pm }(k)$ being the quark and antiquark Fermi-Dirac distributions
\begin{equation}
n_{\pm }(k)=\frac{1}{e^{\left[ E(k)\pm \mu \right] /T}+1}.
\end{equation}
While $I(T,\mu)$ is finite, $I_{quad}({M}^{2})$ is divergent. The divergent
integral is an element of the systematization adopted in the
calculational strategy \cite{B_EJPC} which we use in the present work. In 4D
one-loop calculations, within the context of theories and models where the
divergence degree is not higher than cubic, which is the case of the NJL
model, besides $I_{quad}(M^{2})$ there also appears the basic logarithmically
divergent object
\begin{equation}
I_{log}(M^{2})=\int \frac{d^{4}k}{(2\pi )^{4}}\frac{1}{(k^{2}-M^{2})^{2}}.
\end{equation}
In the NJL model, $I_{quad}(M^{2})$ is related to the constituent quark mass,
as shown in Eq. (\ref{Gap_M}), while $I_{log}(M^{2})$ is related to the pion
decay constant in vacuum -- see Eq.~(\ref{fpi}).

The traditional procedure to treat the divergent amplitudes
in nonrenormalizable models is the introduction of a regularization
distribution in all loop integrals in order to render them convergent.
In the context of the NJL model, with rare exceptions -- such as in Refs.~
\cite{Blaschke,Ruivo,Weise_fin_int} -- the \emph{finite} integral
$I(T,\mu _{B})$ containing the Fermi-Dirac distributions is also
regularized. Due to the nonrenormalizable character of the model,
the regularization introduced in this way cannot be removed along
the calculations and therefore the predictions become associated with
the chosen regularization. In practice, the amplitudes
become functions of the regularization distribution parameters and
consequently the energy-momentum dependence of the amplitudes, as emerging
from the Feynman rules, are modified. As it is well-known, this leads
to violation of fundamental symmetries like those associated to
space-time homogeneity or to the introduction of non-physical
thresholds that break unitarity.

Having this in mind, an alternative strategy to handle the divergences
typical of the perturbative calculations in QFT has been
developed in Ref.~\cite{thesis}. The main idea is to avoid the critical step
of the explicit evaluation of a divergent integral. The finite and
divergent parts are separated by using a convenient representation
for the propagators, in that all the dependence on the physical momenta
is located in finite integrals which are not affected by regularization,
thus avoiding the contamination with non-physical behaviors. The divergent
parts obtained in this way are reduced to only a small number of basic
objects after the adoption of a set of properties for purely divergent
integrals, denominated consistency relations (CR), which are dictated by
the requirements of symmetry preservation and elimination of
ambiguities~\cite{Bc}. The remaining
divergent objects need not be solved explicitly; in renormalizable theories
they are completely absorbed in renormalization of physical parameters, and
in nonrenormalizable models they can be directly adjusted to physical
quantities.

There are many advantages in the adoption of the procedure just described,
but two of them can be immediately noted: the treatment for the divergences
is universal and the energy-momentum dependence of the amplitudes is not
modified. Besides, a connection with traditional procedures can be made.
In particular, all results can be mapped to the ones obtained within the
context of dimensional regularization (DR), in situations where this
method applies. On the other hand, the connection with regularizations
based on regularizing distributions requires some care in the case of
nonrenormalizable models, due to the fact that the effects of the
regularization on finite integrals are removed in our strategy.
In this sense it is important to say that the prescription described
above, which has been applied in many problems involving divergences
in QFT \cite{impl-diff}, represents a particular prescription to handle
the amplitudes having divergences. Therefore, it should be considered
a procedure to make predictions with nonrenormalizable models in the
same way as with any other regularization prescription. However, the
important point to be noted is that, by construction many of the
problems intrinsic to the traditional methods have been circumvented,
like violations of symmetry relations among Green's functions,
ambiguities, non-physical thresholds and scale invariance breaking.
The convenience or usefulness of the method for the treatment of
nonrenormalizable models should be judged by the consistency
of its phenomenological implications.

At one loop order, only two divergent integrals, $I_{quad}(M^{2})$
and $I_{log}(M^{2})$ need be eliminated by some phenomenological
adjustment. We first note that they are not independent quantities,
they obey two scaling relations \cite{B-PRD-2005}
\begin{eqnarray}
\hspace{-0.5cm}I_{quad}(M^{2}) &=&I_{quad}(\lambda ^{2})+\left(
M^{2}-\lambda ^{2}\right) \,I_{log}(\lambda ^{2})   \nonumber \\
&+&\frac{i}{(4\pi )^{2}}\left[ M^{2}-\lambda ^{2}-M^{2}\log \left( \frac{
M^{2}}{\lambda ^{2}}\right) \right] , \label{SR_1} \\
\hspace{-0.5cm}I_{log}(M^{2}) &=&I_{log}(\lambda ^{2})-\frac{i}{(4\pi )^{2}}
\log \left( \frac{M^{2}}{\lambda ^{2}}\right) ,  \label{SR_2}
\end{eqnarray}
where $\lambda ^{2}$ is the arbitrary mass scale adopted in the separation
of finite and divergent parts. These relations allow us to relate the basic
divergent quantities in two different mass scales, as we shall see in a
moment. The imposition of scale independence implies in two properties for
these objects
\begin{eqnarray}
\frac{\partial I_{quad}(\lambda ^{2})}{\partial \lambda ^{2}}
&=&I_{log}\left( \lambda ^{2}\right) ,  \label{rel_1} \\
\frac{\partial I_{log}\left( \lambda ^{2}\right) }{\partial \lambda ^{2}}
&=&-\frac{i}{(4\pi )^{2}}\frac{1}{\lambda ^{2}}.  \label{rel_2}
\end{eqnarray}
It is important to note that only finite quantities have
been differentiated. The above properties work like requirements that must
be imposed in regularizations in order to maintain the scaling properties of
the amplitudes. The above referred properties can be understood within the
context of regularizations. For this purpose, we assume that each integral
is regularized through an unspecified distribution $f\left( k/\Lambda \right)
$, where $\Lambda $ is a parameter with the dimensions of momentum such that
\begin{eqnarray}
&&I_{quad}(M^{2})=\int \frac{d^{4}k}{(2\pi )^{4}}\frac{f(k/\Lambda )}{
k^{2}-M^{2}},  \label{Iquad} \\
&&I_{log}(M^{2})=\int \frac{d^{4}k}{(2\pi )^{4}}\frac{f(k/\Lambda )}{
(k^{2}-M^{2})^{2}}.  \label{Ilog}
\end{eqnarray}
Differentiating $I_{quad}(M^{2})$, relation (\ref{rel_1}) follows
immediately. By differentiating (\ref{Ilog}), on the other hand,
we obtain
\begin{equation}
\frac{\partial I_{log}\left( M^{2}\right) }{\partial M^{2}}=2\int \frac{
d^{4}k}{(2\pi )^{4}}\frac{f(k/\Lambda )}{(k^{2}-M^{2})^{3}} .
\end{equation}
Since the integral is finite and assuming the existence of the limit
$_{\Lambda \rightarrow \infty }^{lim}f(k/\Lambda )=1$, one can extract the
regulating distribution and perform the integration to obtain (\ref{rel_2}).
This last point illustrates the basic difference between our prescription and the
traditional ones.

The fact that the parameter $\Lambda$ plays the role of a cutoff can be made
clear even in the case of an implicit regularization. That is, it is
possible to make the $\Lambda $ dependence explicit without
specifying the regulating function. This can be done by defining the
dimensionless ratio $M_{\Lambda }=M/\Lambda $ and writing $I_{quad}(M^{2})$
and $I_{log}(M^{2})$ in terms of dimensionless integrals
$J_{quad}(M_{\Lambda }^{2})$ and $J_{log}(M_{\Lambda }^{2})$ as
\begin{eqnarray}
I_{quad}(M^{2}) &=&\Lambda ^{2}J_{quad}(M_{\Lambda }^{2}),  \label{defJq} \\
I_{log}(M^{2}) &=&J_{log}(M_{\Lambda }^{2}),  \label{defJl}
\end{eqnarray}
with
\begin{eqnarray}
J_{quad}(M_{\Lambda }^{2}) &=&\int \frac{d^{4}u}{(2\pi )^{4}}\frac{f(u)}{
u^{2}-M_{\Lambda }^{2}},  \label{Jquad} \\
J_{log}(M_{\Lambda }^{2}) &=&\int \frac{d^{4}u}{(2\pi )^{4}}\frac{f(u)}{
(u^{2}-M_{\Lambda }^{2})^{2}},  \label{Jlog}
\end{eqnarray}
where the integration variable $u$ is dimensionless. The integrals $
J_{quad}(M_{\Lambda }^{2})$ and$\ J_{log}(M_{\Lambda }^{2})$ obey scaling
relations completely similar to that (\ref{SR_1}) and (\ref{SR_2}), as
they should. A general parametrization for these integrals, obeying the
properties (\ref{rel_1}) and (\ref{rel_2}), can be constructed and written as
\begin{eqnarray}
i\left( 4\pi \right) ^{2}J_{log}(M_{\Lambda }^{2}) &=&\ln \left( M_{\Lambda
}^{2}\right) +c_{1}, \\
i\left( 4\pi \right) ^{2}J_{quad}(M_{\Lambda }^{2}) &=&M_{\Lambda }^{2}\ln
\left( M_{\Lambda }^{2}\right) -M_{\Lambda }^{2}+c_{2}.
\end{eqnarray}
It should be clear that if a regularization distribution does not furnish the
above general form, the scaling properties of the amplitudes can be broken.
In principle, different (consistent) regularizations should differ only by
the values of the constants $c_{1}$ and $\ c_{2}$.

After this review on the general procedure to handle the divergences, let us
turn our attention to its application to the NJL model. First, we have that
the vacuum quark condensate is related to the scalar one-point function,
which can be written as
\begin{equation}
\langle \overline{\psi }\psi \rangle _{0}=-12M_{0}\,iI_{quad}(M_{0}^{2}),
\label{cond}
\end{equation}
where $M_{0}$ is the mass obtained by solving the gap equation in vacuum. On
the other hand, the pion decay constant $f_{\pi }$ can be related to the
axial-pseudoscalar two-point function, such that
\begin{equation}
f_{\pi }^{2}=-12M_{0}^{2}\,iI_{log}(M_{0}^{2}).  \label{fpi}
\end{equation}%
The point now is that one can express the temperature- and density-dependent
gap equation~(\ref{Gap_M}) for quark mass $M$ in terms of these vacuum
quantities. This is so because in general the scaling relations (\ref{SR_1})
and (\ref{SR_2}) can be used to isolate vacuum contributions from temperature
- and density-dependent contributions in expressions for physical quantities,
that is, it is possible to write $I_{quad}(M^{2})$ and$\ I_{log}(M^{2})$ in
terms of $I_{quad}(M_{0}^{2})$ and $I_{log}(M_{0}^{2})$. Using the
scaling relation in (\ref{Gap_M}), we get
\begin{eqnarray}
M &=&48\,G\,M\Biggl\{-\frac{\langle \overline{\psi }\psi \rangle _{0}}{
12M_{0}}-(M^{2}-M_{0}^{2})\frac{f_{\pi }^{2}}{12M_{0}^{2}}  \nonumber \\
&-&\frac{1}{(4\pi )^{2}}\left[ M^{2}-M_{0}^{2}-M^{2}\log \left( \frac{M^{2}}{
M_{0}^{2}}\right) \right] \hspace{0cm}-I(T,\mu )\Biggr\},  \nonumber \\
\label{gap_T_mu}
\end{eqnarray}
where we have made use of Eqs. (\ref{cond}) and (\ref{fpi}). The
corresponding expression for the case $m_{0}\neq 0$ is a little more
complicated than the one shown in Eq. (\ref{gap_T_mu}), since the expression
for $f_{\pi }^{2}$ contains, in addition to $I_{log}$, a finite integral
involving the pion mass.

\begin{figure}[tbh]
\includegraphics[scale=0.525]{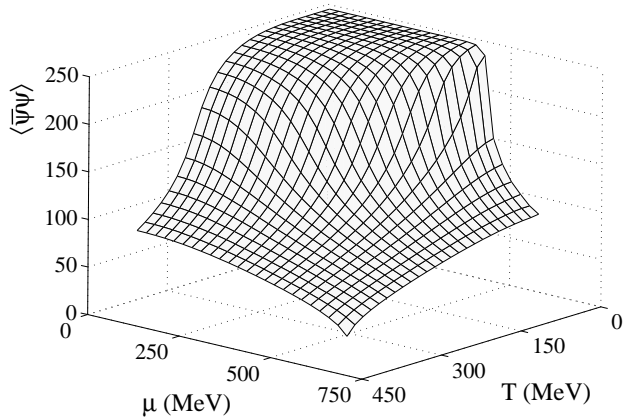}
\caption{One flavor quark condensate $\langle \overline{\protect\psi }
\protect\psi \rangle $ as function of temperature $T$ and quark chemical
potential $\protect\mu $.}
\label{cond_T_mu}
\end{figure}

For illustrative purposes, we present in Fig.~\ref{cond_T_mu} the quark
condensate as a function of temperature $T$ and quark chemical potential $
\mu $ -- the results here are obtained solving the gap equation with $m_0 =
5.5$~MeV, $G = 4.9 \times 10^{-6}$~MeV$^{-2}$ and the fitting mass $M_0=312$
~MeV. As seen in Fig.~\ref{cond_T_mu}, this implicit regularization scheme gives
the expected reduction of the quark condensate in medium. In addition, this
reduction is in qualitative agreement with the result obtained with
traditional cutoff regularization schemes - see for example Fig.~26 of Ref.~
\cite{VW}. In the chiral limit, the order of the phase transition is first
order. It is also worth mentioning that with these same parameters one is
able to obtain good values for the $\pi$- and $\sigma$-meson masses~\cite{BDK}.

Of course, the point of the exercise here was not only to show that this
different way of handling the divergent integrals gives results in accord
with the expected phenomenology. What should be noted here is that once we
have eliminated $I_{quad}(M_{0}^{2})$ and $I_{log}(M_{0}^{2})$ in favor of $%
\langle \bar{\psi}\psi \rangle _{0}$ and $f_{\pi }$ we have kept $\Lambda $
implicitly fixed. If one wants to change $\Lambda $ without changing the
low-energy results $G$ has to run with $\Lambda $, as in Ref.~\cite{Casal}.
The point here is that one can implement this using the scaling relations
(\ref{SR_1}) and (\ref{SR_2}), which are independent of an explicit
regularization. This will be done in the next Section.

\section{Running of the coupling}

\label{sec:running}

Fixing $f_{\pi }=93$~MeV and using the expression of $f_{\pi }$ in terms of $%
I_{log}$ at the scales $(M,\Lambda )$ and $(M_{0},\Lambda _{0})$, one
obtains $M=M(\Lambda )$ as the solution of the transcendental equation
\begin{equation}
M^{2}=M_{0}^{2}\,\frac{\Lambda ^{2}}{\Lambda _{0}^{2}}\,\exp \Bigg[\frac{
4\pi ^{2}}{3}\frac{f_{\pi }^{2}}{M^{2}}\,\bigg(\frac{M^{2}}{M_{0}^{2}}-1
\bigg)\,\Bigg].
\label{MxLambda}
\end{equation}
Using the value of $\langle \bar{\psi}\psi \rangle _{0}$ at the reference
scale $(M_{0},\Lambda _{0})$, and using the scaling relation~(\ref
{SR_1}) to obtain $I_{quad}$ at $(M,\Lambda )$ in the gap equation, one
obtains $G=G(\Lambda )$ as
\begin{eqnarray}
\left[48 \Lambda^2 G(\Lambda) \right]^{-1} &=&
\frac{-\langle\bar{\psi}\psi\rangle_0}{12M_0\Lambda_0^2} -
\Bigg(\frac{M^2}{\Lambda^2} - \frac{M_0^2}{\Lambda_0^2}\Bigg)
\frac{f_\pi^2}{12M_0^2}\nonumber\\
&-& \frac{1}{(4\pi)^2}\Bigg[\frac{M^2}{\Lambda^2} -
\frac{M_0^2}{\Lambda_0^2} - \frac{M^2}{\Lambda^2}
\log\Bigg(\frac{M^2\Lambda_0^2}{\Lambda^2M_0^2} \Bigg) \Bigg],
\end{eqnarray}
where it is understood that $M=M(\Lambda )$, as given by
Eq.~(\ref{MxLambda}).

This is our main result in the present paper. It is an interesting result,
in that it was obtained without specifying any explicit regularization, only
very general scaling relations of the divergent integrals were used. In this
sense, the result seems very general and robust. Once a temperature and
density dependence for $\Lambda$ is specified, $G$ becomes also temperature
and density dependent.

\vspace{0.5cm}
\begin{figure}[htb]
\centerline{\includegraphics[scale=0.325]{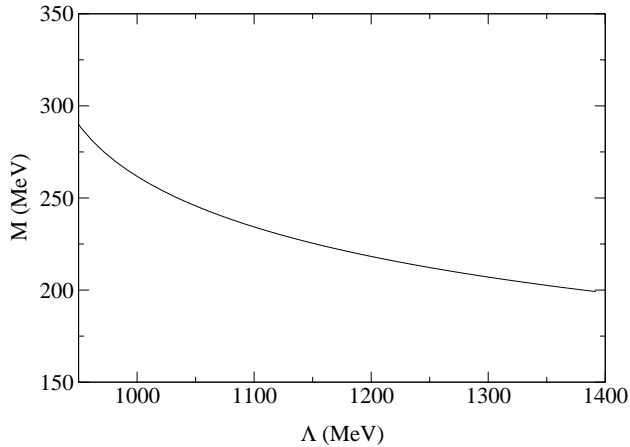}}
\caption{Constituent quark mass M as a function of $\Lambda$.}
\label{mass_Lambda}
\end{figure}

\begin{figure}[htb]
\centerline{\includegraphics[scale=0.325]{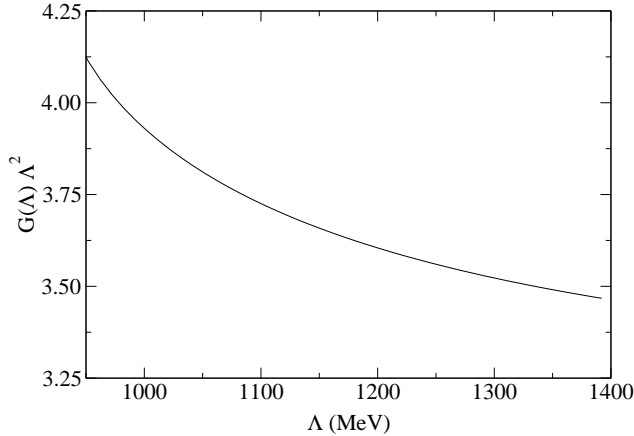}}
\caption{Running of the coupling $G(\Lambda)$.}
\label{GL2_Lambda}
\end{figure}

In Fig.~\ref{mass_Lambda} we present the numerical results for the $\Lambda$
dependence of $M$, as obtained from the solution of Eq.~(\ref{MxLambda})
after solving the gap equation. The constituent mass decreases as $\Lambda$
is increased. If $G$ were kept fixed, $M$ would increase of course. But $G$
decreases with $\Lambda$, as shown in Fig.~\ref{GL2_Lambda}, and the net
effect is that $M$ decreases. These results for $M=M(\Lambda)$ and $
G=G(\Lambda)$ are in qualitative agreement with the results of Ref.~\cite
{Casal}.

\vspace{0.5cm}

\vspace{0.5cm}

The specification of a density and temperature dependence for $\Lambda$ might
seem a little arbitrary. But this needs not be the case, because the entire line
of arguments can be turned around, instead of fixing~$f_\pi$ at some value,
one could \emph{postulate} in a physically motivated way a running behavior
for $G = G(\Lambda)$ and work backwards. What would change in this case?
First, using the postulated $G(\Lambda)$ in the gap equation, one would
obtain a corresponding $M(\Lambda)$. When this $M(\Lambda)$ is replaced in
the expression for~$f_{\pi}$, in general~$f_{\pi}$ will be also $\Lambda$
dependent, but this $\Lambda$ dependence would be very weak since the
integral for~$f_{\pi}$ is only logarithmically divergent. In this way, the
chiral physics in vacuum would be maintained. This is very interesting,
since one could use the predicted density and temperature running of the
QCD coupling constant for $G$ and in this way fit the density and temperature
dependence of $\Lambda$. In this way, one would be modeling, admittedly in
a crude way, the asymptotic freedom of QCD in the NJL model.

In closing this Section we reiterate that the purpose for making $G$ to run
with $\Lambda$ is to extend the applicability of the model to high densities
and temperatures. At high densities and temperatures, high momentum
components are present in the system and a cutoff of the order of the chiral
symmetry breaking scale invalidates the use of the model in such situations.
In order to illustrate the use of the extension in practice we consider in
the next Section color superconductivity in high-density quark matter.

\section{Color superconductivity}

\label{sec:CSC}

For reviews and a comprehensive list of references on the subject of color
superconductivity see Refs.~\cite{RW,Alf,Shov,Nard,Rch,Bub,Huang,Scha}.
Since the aim here is to illustrate the formalism we simplify matters by
using the same Lagrangian density as used above, although for obtaining a
better phenomenological description of both chiral symmetry breaking and
color superconductivity a more general four-fermion Lagrangian should be
used~\cite{skp,steiner,NA}. At the one-loop level the self-consistent
equation for the superconducting gap is given by (we use the same letter $G$
to denote both the diquark coupling here and the quark-antiquark coupling in
the last section)
\begin{equation}
1=16\,G\int \frac{d^{4}k}{(2\pi )^{4}}\frac{i}{
k_{0}^{2}-\left( k+\mu \right) ^{2}-\Delta ^{2}}+(\mu \rightarrow -\mu ).
\label{gap-super}
\end{equation}
The integrals above are divergent. The application of the implicit
regularization scheme to this problem proceeds as follows~\cite{FDKB}.
Instead of introducing a cutoff in the integrals, the integrands are assumed
to be implicit regularized and then manipulated in a way divergences are
isolated in $\mu $-independent divergent integrals. These divergent
integrals can be related to the divergent integrals $I_{quad}$ and $I_{log}$
of the problem of chiral symmetry breaking in vacuum though the use of the
scaling relations discussed above. From the manipulation of the integrand in
Eq.~(\ref{gap-super}) also result finite integrals, and these are integrated
without imposing any restriction to their integrands. Initially we consider $%
T=0$. In this case the equation for the superconducting gap is given by
\begin{eqnarray}
1 &=&2\,G\left( \Lambda \right) \Lambda ^{2}\,\left\{ -\frac{4}{3}\,\frac{%
\langle \overline{\psi }\psi \rangle _{0}}{M_{0}\,\Lambda _{0}^{2}}-\frac{4}{%
3}\,\left( \frac{\Delta ^{2}}{\Lambda ^{2}}-\frac{M_{0}^{2}}{\Lambda _{0}^{2}%
}\right) \frac{f_{\pi }^{2}}{M_{0}^{2}}\right.  \nonumber \\
&-&\frac{1}{\pi ^{2}}\,\left[ \frac{\Delta ^{2}}{\Lambda ^{2}}-\frac{%
M_{0}^{2}}{\Lambda _{0}^{2}}-\frac{\Delta ^{2}}{\Lambda ^{2}}\log \left(
\frac{\Delta ^{2}\Lambda _{0}^{2}}{\Lambda ^{2}M_{0}^{2}}{\ }\right) +2\frac{%
\mu ^{2}}{\Lambda ^{2}}\right]  \nonumber \\
&+&\left. \frac{\mu ^{2}}{\Lambda ^{2}}\left[ \frac{8}{3}\frac{f_{\pi }^{2}}{%
M_{0}^{2}}-\frac{2}{\pi ^{2}}\,\log \left( \frac{\Delta ^{2}\Lambda _{0}^{2}%
}{\Lambda ^{2}M_{0}^{2}}\right) \right] \right\} ,  \label{gap2}
\end{eqnarray}%
where we used the manipulations of the integrand in Eq.~(\ref{gap-super}) as
explained above and shown explicitly in Ref.~\cite{FDKB}.

\begin{figure}[htb]
\centerline{\includegraphics[scale=0.355]{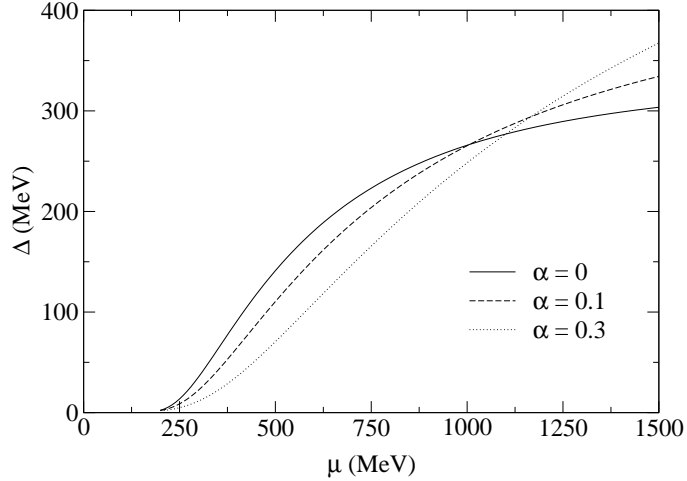}}
\caption{Zero temperature superconducting gap as a function of quark
chemical potential for three different values of $\protect\alpha$ in Eq.~(%
\protect\ref{Lambda_mu}).}
\label{alphas}
\end{figure}

One very interesting result of the application of the implicit
regularization to the problem of color superconductivity is that, as shown
with greater detail in Ref.~\cite{FDKB}, the superconducting gap as a
function of $\mu$ does not vanish at $\mu \simeq \Lambda$, as happens with
the traditional cutoff schemes. The numerical result for $\Delta$ is shown
by the solid line in Fig.~\ref{alphas}. This is interesting because a
nonvanishing gap at high quark densities is predicted by QCD~\cite{son} -
see also the reviews in Ref.~\cite{RW,Alf,Shov,Nard,Rch,Bub,Huang,Scha}. Of
course, if the implicit regularization scale $\Lambda$ is kept fixed, the
use of the model at high densities is questionable. However, we are able to
make explicit the $\Lambda$-dependence in the gap equation (and in other
physical quantities as well) by extracting the $\Lambda$ dependence from the
implicit regularization function. In this way, one can very easily extend
the applicability of the model to larger values of $\mu$ by allowing a
running $G(\Lambda)$ and a $\mu$ dependence for $\Lambda$.

\vspace{0.5cm}

\vspace{0.5cm}

\begin{figure}[ht]
\centerline{\includegraphics[scale=0.355]{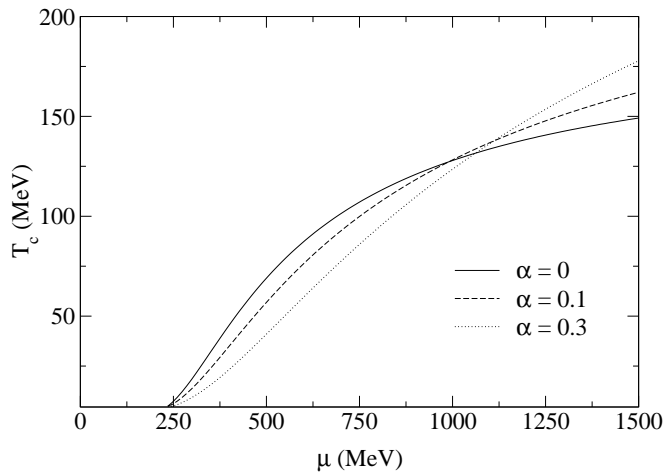}}
\caption{Critical temperature for the superconducting gap as a function of 
$\mu$ for different values of $\alpha$ in Eq.~(\ref{Lambda_mu}). }
\label{Tc}
\end{figure}

For the density dependence of $\Lambda$, as said above, there is a great deal of
arbitrariness. As discussed, one could fix this dependence, for instance,
by matching the density dependence of $G(\Lambda)\Lambda^2$ with the prediction
of perturbative QCD for the running of the QCD coupling constant $\alpha_s$ at
high densities -- the one loop prediction is that the coupling decreases
logarithmically with $\mu$ for large values of $\mu$. However, for our
purposes here of showing the qualitative results only, we use the simple
formula for $\mu \ge \mu_0 = 235$~MeV,
\begin{eqnarray}
\Lambda=\Lambda_0 \left[1 + \alpha \, \log\left(\frac{\mu}{\mu_0}\right)%
\right] ,  \label{Lambda_mu}
\end{eqnarray}
where $\alpha$ is a constant. When this is used into the expression for $%
G(\Lambda)\Lambda^2$, one obtains that $G(\Lambda)\Lambda^2$ decreases
logarithmically with $\mu$, mocking up in a rather crude way the prediction
of perturbative QCD for the running of $\alpha_s$ with $\mu$. In Fig.~\ref%
{alphas} we show the numerical results $\Delta$ as a function $\mu$ for
different values of $\alpha$. We use for the diquark pairing strength the
value $G = 3.1$~GeV$^{-2}$. The other parameters are $M_0 = 312$~MeV and $%
\Lambda_0 = 932$~MeV, which are the values obtained in vacuum for $f_\pi =
93 $~MeV and $\langle\bar{\psi}\psi\rangle_0 = ( -~250.\,\mathrm{{MeV})^3}$.
The results show the expected behavior of the gap growing faster with $\mu$
as $\Lambda$ increases with $\mu$, i.e. as $\alpha$ increases.

For completeness, we include the effects of temperature. We calculate the
critical temperature $T_{c}$ above which $\Delta =0$, for different values
of $\mu $. We rewrite Eq.~(\ref{gap-super}) for $\Delta =0$ including the
effect of temperature
\begin{eqnarray}
1 &=&16\,G\int \frac{d^{4}k}{\left( 2\pi \right) ^{4}}\frac{i}{%
k_{0}^{2}-\left( k+\mu \right) ^{2}}\tanh \left( \frac{\beta }{2}|k+\mu
|\right)  \nonumber \\
&+&(\mu \rightarrow -\mu ),
\end{eqnarray}
where
\begin{equation}
\tanh \left( \frac{\beta }{2}\left\vert k\pm \mu \right\vert \right)
=1-2n_{\pm }^{\prime }\left( k\right) ,
\end{equation}
with
\begin{equation}
n_{\pm }^{\prime }\left( k\right) =\frac{1}{e^{\beta \left\vert k\pm \mu
\right\vert }+1}.
\end{equation}
Using the same manipulations to isolate the divergent integrals as explained
before and after some algebraic effort we can write
\begin{eqnarray}
\frac{1}{4G} &=&\Lambda ^{2}\Biggl\{-\frac{2}{3}\frac{\langle \bar{\psi}\psi
\rangle _{0}}{M_{0}\Lambda _{0}^{2}}+\frac{2}{3}\left( \frac{M_{0}^{2}}{%
\Lambda _{0}^{2}}+\frac{2\mu ^{2}}{\Lambda ^{2}}\right) \frac{f_{\pi }^{2}}{%
M_{0}^{2}}  \nonumber \\
&+&\frac{1}{2\pi ^{2}}\left[ \frac{M_{0}^{2}}{\Lambda _{0}^{2}}-\frac{3\mu
^{2}}{\Lambda ^{2}}-\frac{2\mu ^{2}}{\Lambda ^{2}}\ln \left( \frac{\mu ^{2}}{%
\Lambda ^{2}}\frac{\Lambda _{0}^{2}}{M_{0}^{2}}\right) \right] \Biggr\}
\nonumber \\
&+&Q(\mu ,T)+Q(-\mu ,T),
\end{eqnarray}
where
\begin{eqnarray}
Q(\mu ,T) &=&-\frac{2}{\pi ^{2}}\int_{0}^{\infty }dk\frac{k^{2}}{\left(
k^{2}+\mu ^{2}\right) ^{1/2}}\left[ n_{+}^{\prime }\left( k\right)
+n_{-}^{\prime }\left( k\right) \right]  \nonumber \\
&+&\frac{2\mu }{\pi ^{2}}\int_{0}^{\infty }dk\frac{k^{3}}{\left( k^{2}+\mu
^{2}\right) ^{3/2}}\left[ n_{+}^{\prime }\left( k\right) -n_{-}^{\prime
}\left( k\right) \right]  \nonumber \\
&-&\frac{3\mu ^{2}}{\pi ^{2}}\int_{0}^{\infty }dk\frac{k^{4}}{\left(
k^{2}+\mu ^{2}\right) ^{5/2}}\left[ n_{+}^{\prime }\left( k\right)
+n_{-}^{\prime }\left( k\right) \right]  \nonumber \\
&+&\frac{1}{2\pi ^{2}}\int_{0}^{\infty }dk\,k^{2}\,I(k)\,\tanh \left( \frac{%
\beta }{2}\left\vert k+\mu \right\vert \right) ,
\end{eqnarray}
with
\[
I(k)=\frac{2k^{5}+5k^{2}\mu ^{2}(k-\mu )-2\mu ^{5}-2(k^{2}+\mu ^{2})^{5/2}}{%
(k^{2}+\mu ^{2})^{5/2}(k-\mu )}.
\]

In Fig.(\ref{Tc}) we plot the critical temperature as function of $\mu$. In
the solid line we have kept the coupling $G$ fixed and in the dashed and
dotted lines the coupling is a function of $\mu$ through the $\mu$
dependence of $\Lambda$ as in Eq.~(\ref{Lambda_mu}). As seen, the critical
temperature increases with $\mu$, and this increase is faster as $\alpha$
increases for large $\mu$. The values of $T_c$ obtained within this scheme
seem a little larger than obtained with cutoff regularization.

\section{Conclusions and perspectives}

\label{sec:concl}

We have considered an extension of the NJL model to high densities and
temperatures as proposed in Ref.~\cite{Casal} and used an alternative strategy
to handle ultraviolet divergences. This extension is implemented by allowing
the regularization scale $\Lambda $ to increase at high densities with the
simultaneous decrease of the coupling $G$. Making use of the scaling
relations of Eqs.~(\ref{SR_1}) and (\ref{SR_2}), and the definitions in
Eqs.~(\ref{defJq}) and (\ref{defJl}), the two one-loop divergent integrals $%
I_{quad}$ and $I_{log}$ at scales $\Lambda _{0}$ and $M_{0}$ can be related
to the same $I_{quad}$ and $I_{log}$ at different scales $\Lambda $ and $M$.

Although our numerical results are in qualitative agreement with the ones of 
Casalbuoni et al.~\cite{Casal} for the $\mu $ dependence of the
superconducting gap $\Delta $, we believe the present approach is different
from their results in several aspects. Perhaps the most important one is the
fact that no specific regularization distribution was used to obtain the
running of $G$ with $\Lambda $. Proceeding this way it is possible to
circumvent many troubles commonly found when traditional regularization
schemes are used, like those based on a three- or four-momentum cutoff, or
Pauli-Villars and proper-time regularizations, lead in general to global and
gauge symmetry violations, and to the breaking of causality and unitarity.
Although in some circumstances such aspects might not be of crucial importance,
in many others, like in hadronic correlation functions, remarkable differences
appear~\cite{FariasHad}. Among them, only physical thresholds
(independent of $\Lambda $) are present allowing thus the preservation
of unitarity. In addition, it is easy to show that causality is also
preserved by checking that the amplitudes obey the correct dispersion
relations. Gauge symmetry is also preserved and the ambiguities are
eliminated at the one-loop level, as shown in Ref.~\cite{BN} in the context
of the gauged NJL model. A complete discussion of these matters are
discussed in detail in a separate publication~\cite{BDK}.

Another interesting aspect of the present approach is that it can be used
with heavy quarks. The scaling relations involve the ratio of the quark mass
to the implicit regulation scale and naturally take into account
short-distance effects as the quark mass increases. This is very important
in connection with studies of heavy-quark bound states in highly excited
quark matter.

\vspace{1.0cm}

\textbf{Acknowledgments}: Work partially supported by the Brazilian agencies
CAPES, CNPq, FAPERJ and FAPESP.

\end{document}